\documentstyle[12pt]{article}
\input{epsf}
\newcommand{\be}{\begin{eqnarray}}
\newcommand{\ee}{\end{eqnarray}}
\newcommand{\bi}{\bibitem}

\begin{document}
{\hbox to\hsize{January, 1999 \hfill TAC-1999-002}
\begin{center}
{\Large \bf{ Geometry and Statistics of Cosmic Microwave Polarization.} }\\
\bigskip
{\bf A.D. Dolgov
\footnote{Also: ITEP, Bol. Cheremushkinskaya 25, Moscow 113259, Russia.},
 A.G. Doroshkevich}\\
{\it {Theoretical Astrophysics Center\\
 Juliane Maries Vej 30, DK-2100, Copenhagen, Denmark}\\
{\bf {D.I. Novikov
\footnote{ Also: Astro - Space Center of Lebedev Physical Institute,
Profsoyuznaya 84/32, Moscow, Russia.}}}\\
{\it {Dept. of Physics and Astronomy, University of Kansas\\
  Lawrence, Kansas, 66045, USA}}\\
{\bf{I.D. Novikov}}
\footnote{Also: Astro - Space Center of Lebedev Physical Institute,
Profsoyuznaya 84/32, Moscow, Russia; University Observatory,
Juliane Maries Vej 30, DK-2100, Copenhagen;
NORDITA, Blegdamsvej 17, DK-2100, Copenhagen, Denmark.}}
\\
{\it {Theoretical Astrophysics Center\\
 Juliane Maries Vej 30, DK-2100, Copenhagen, Denmark}}
\date{\today}
\end{center}
\begin{abstract}

Geometrical and statistical properties of polarization of CMB are analyzed.
Singular points of the vector field which describes CMB polarization are
found and classified. Statistical distribution of the singularities is studied.
A possible signature of tensor perturbations in CMB polarization is discussed. 
For a further analysis of CMB statistics Minkowski functionals are used, which 
present a technically simple method to search for deviations from a Gaussian 
distribution.

\end{abstract}


\section{Introduction}
 
Future high precision measurements of anisotropies of CMB are potentially
very powerful tools for cosmological studies. They may present 
accurate data on the values of the basic cosmological parameters,
as well as information about structure formation at an early linear stage.
There is a very rich literature on this subject; for a review and a partial
list of references see e.g. \cite{rev1}-\cite{rev6}. In addition to
angular variations of CMB temperature, primordial density fluctuations
induce also, as a secondary effect, a polarization of the 
radiation. A good introduction to the subject can be found in the
pioneering papers~\cite{bp}-\cite{be}, while a more up-to-date development
is given in refs.~\cite{gw1}-\cite{zs}. 
Our approach here is slightly different from that in the last papers and, 
though the results mostly coincide or agree, there are
some essential differences which we will specify in Secs. 2 and 3 (see
also our earlier paper~\cite{ddnn}).
 
Measurements of the polarization of
CMB will permit to obtain an additional information which may help to resolve
ambiguities in extracting cosmological parameters from the observational 
data (for a recent discussion of these ambiguities see e.g. ref.~\cite{efbo}).
In particular, polarization is quite sensitive to the presence of tensor
perturbations (gravitational waves) and, as was shown in 
refs.~\cite{gw1}-\cite{gw5} a deviation from zero of the so called
pseudoscalar or "magnetic" part of polarization would be unambiguous 
indication of the presence of gravitational waves. In what follows we 
will further elaborate this issue and generalize some of the results of
ref.~\cite{geom2}. 
 
Geometrical properties of polarization field were actively studied
in the recent years~\cite{gw4}-\cite{geom2}. Important features, 
which characterize geometry and topology of polarization, are the types 
of singularities of the vector flux lines tangent to
the  direction of maximum polarization~\cite{ddnn} (for an earlier paper
see ref.~\cite{nn}). This vector is parallel to one of the
eigenvectors of the Stokes matrix and have the magnitude equal to
the magnitude of polarization. The points where this vector vanishes,
so that no direction is determined, are singular points on the polarization
map and the character of the singularity to a large extend determines
the relief of the polarization map, even rather away from the points of
vanishing polarization. The maps simulated in different papers, including
ours (see Fig. 4), demonstrate that the behavior
(topology) of the flux lines of the polarization field at the points where
polarization is non-vanishing, is determined by the type of the singularity
at vanishing polarization. Hence one may make a conclusion about the types
of the singular points studying polarization maps in the regions where 
polarization is measurable. The properties of these singular points are
discussed in detail in Secs. 4 and 5.
 
Statistical properties of the anisotropy of CMB temperature and CMB 
polarization are of primary importance for an understanding of their 
origin. At the present day only two mechanisms of generation of 
primordial density perturbations are known: inflation and topological 
defects. Simplest inflationary scenarios predict Gaussian 
perturbations that results in the Gaussianity of the CMB temperature 
fluctuations and polarization at the surface of last 
scattering. Tests of a Gaussian nature of the temperature fluctuations 
of CMB, together with a study of its polarization, are important probes 
of inflation. A possible method of testing a Gaussianity of CMB is a study
of statistics of the singular points discussed in Sec. 4.4.
The same statistical tests may also be useful for discrimination of signal 
from noise because Gaussian distributions of both temperature fluctuation 
and polarization generated by noise are rather improbable. 

Statistics of global geometrical properties of CMB maps, both for 
temperature fluctuations and polarization, can be efficiently studied with 
Minkowski Functionals (MF)~\cite{min,had}.  
Applications of this approach to the temperature 
fluctuations and to polarization has been discussed in 
refs.~\cite{wk}-\cite{nfs} and refs.~\cite{nn,ddnn}, respectively.
In a recent paper~\cite{psf} genus statistic of simulated polarization 
maps was analyzed. In Sec. 6 we apply all three MF's, which are known for
two dimensional maps, for the study of statistical properties of CMB 
polarization.
 
This paper is organized as follow: 
The basic concepts and notations are introduced in Sec. 2. 
In Sec. 3 a local description of polarization is compared with a nonlocal one. 
In Secs. 4 and 5 the classification of singular points is discussed and 
various approaches to this problem are compared. The properties of Minkowski 
Functionals are considered in Sec. 6. We finish with Sec. 7 where a short 
discussion of results can be found.

\section{Generalities}

Here we will describe some general properties of CMB polarization and present
the necessary formalism.
This section has an introductory character and to some extend is
described in the literature. We make a simplifying assumption that the
relevant angular scales are sufficiently small, so that  
the corresponding part on the sky is almost flat.
In this approximation the polarization field on the sky can be considered 
as two dimensional field on flat $(x,y)$-plane. 
The photon polarization is described by the second rank tensor
$a_{ij}$ in the plane
perpendicular to the photon propagation. By definition this tensor is
traceless, because the trace part, proportional to unit matrix,
corresponds to zero polarization and
can be absorbed in the total intensity of radiation.
It is convenient to expand this
tensor in terms of the Pauli matrices $\sigma_\alpha$, $\alpha =1,2,3$,
which form a complete system in $2 \times 2$ traceless matrix space:
\be
\mathbf{a} = \xi_\alpha  \sigma_\alpha
\label{asigma}
\ee
The parameter $\xi_2$ is equal to the amplitude of circular polarization,
which is not generated by Thomson scattering (due to parity conservation),
so that it is assumed usually that  $\xi_2 =0$. In this case the matrix
$\mathbf{a}$ is symmetric and is determined by two functions:
\be
\mathbf{a} =
\left( \begin{array}{cc}
Q & U \\
U & - Q
\end{array}
\right)
\label{aqu}
\ee
The functions $Q$ and $U$ depend upon the coordinate frame; they are
components of the tensor $a_{ij}$ and obey the corresponding tensor
transformation law:
\be
a_{ij}' = T_i^k T_j^l a_{kl}
\label{aij'}
\ee
where the coordinate transformation is given by $x_i'= T_i^k x_k $. In
particular under rotation of the coordinate system with
\be
\mathbf{T} =
\left( \begin{array}{cc}
 c & s \\
-s & c
\end{array}
\right)
\label{trot}
\ee
where $c=\cos \phi$, $s =\sin \phi$, and $\phi$ is the rotation angle,
the parameters $Q$ and $U$ are transformed as:
\be
Q' = Q \cos 2\phi + U \sin 2\phi  \nonumber \\
U' = -Q \sin 2\phi + U \cos 2\phi
\label{q'u'}
\ee
It is more convenient in many cases to work with invariant quantities or
at least with vector ones, whose direction on the polarization map is easy
to visualize.  There are the following invariants (or what is the same,
scalars) that may be constructed from the second rank tensor.  
First, of course, it is the trace, 
${\rm Tr}\, {\mathbf{a} }= a_{ii}$. In the
considered case it is zero. The second invariant is the determinant of the
matrix $ \mathbf{a}$,
\be
{\rm det} \,\, {\mathbf{a} }
= Q^2 + U^2
\label{deta}
\ee
The maximum magnitude of polarization is given by $\sqrt {Q^2 + U^2}$.
The direction of maximum polarization is determined by one of the
eigenvectors of the matrix $a_{ij}$ (see e.g. ref.~\cite{ddnn}).

These are the well known algebraic invariants which exist in any 
space dimensions. One may construct
two more invariants, using vector operator of differentiation. They
can be chosen as:
\be
S =  \partial_i  \partial_j a_{ij}  \nonumber \\
P =  \epsilon_{kj}  \partial_k  \partial_i a_{ij}
\label{sp}
\ee
where $j=1,2$ and $\partial_j =\partial /\partial\, x^j$. In terms 
of $Q$ and $U$ these invariants are expressed as:
\be
S=  (\partial_1^2 -\partial_2^2) Q + 2 \partial_1 \partial_2 \, U \nonumber \\
P= (\partial_1^2 -\partial_2^2) U - 2 \partial_1 \partial_2 \, Q
\label{spqu}
\ee
The first scalar invariant exists in any space dimension, while the second
pseudo-scalar one exists only in two dimensional space, because of the
presence of antisymmetric pseudo-tensor $\epsilon_{kj}$ (analogous
antisymmetric
tensor in higher dimensions $D$ has $D$ indices).
These quantities $S$ and $P$ coincide, up to a scalar factor,
with the introduced in refs.~\cite{gw1,gw3} $B$ and $E$ fields. 
To our opinion it is
more natural to denote them as $S$ and $P$ to stress their scalar and
pseudo-scalar nature and not as electric and magnetic parts of polarization
because these quantities have nothing to do with vectors. In this sense we
agree with the terminology of ref.~\cite{sp} (see also \cite{gw4,geom1}).

An important feature of the pseudo-scalar $P$ is that it vanishes if  
only  scalar perturbations induces polarization in CMB. In this
case the Stokes matrix can be written in terms of derivatives of one scalar
function:
\be
a_{ij} = \left( 2\partial_i  \partial_j -
          \delta_{ij} \partial_k \partial_k \right) \Psi
\label{apsi}
\ee
It is straightforward to check that indeed $P=0$. We do not
share the opinion and/or terminology of refs.~\cite{gw5,kam} where it is stated
that the corresponding field does not possess a curl.
As has been shown in ref.~\cite{ddnn} this is not true and generically  the 
eigenvectors of the Stokes matrix are not curless. The validity of this
general statement can be verified on simple examples. It means in particular 
that the flux lines of the direction of maximum polarization may have a 
nonzero vorticity in contrast to the statement of refs.~\cite{gw5,kam}.

If tensor perturbations are non-vanishing, the polarization matrix has
the general form determined by two independent functions.
As is well known, an arbitrary three dimensional vector
can be expanded in terms of scalar and vector potentials as
\be
\vec V = {\rm grad} \,\Phi + {\rm curl} \,{\vec A}
\label{vecv}
\ee
In two dimensions an arbitrary vector can be expressed as derivatives of 
a scalar and a pseudoscalar:
\be
 V_j = \partial_j \Phi_1 + \epsilon_{jk} \partial_k \Phi_2
\label{vecv2}
\ee
In direct analogy to that, an arbitrary traceless symmetric
$2\times 2$-matrix can be presented
in terms of scalar and pseudo-scalar potentials as:
\be
a_{ij} = \left( 2\partial_i  \partial_j -
          \delta_{ij} \partial^2 \right) \Psi
+\left( \epsilon_{ik}\partial_k \partial_j +
\epsilon_{jk}\partial_k \partial_i \right) \Phi
\label{ageneral}
\ee
Of course now the pseudo-scalar $P$ defined in eq. (\ref{sp}) does not vanish
and this property permits to observe possible tensor perturbations by
measurement of CMB polarization~\cite{gw1}-\cite{geom2}. If $\Psi =0$ then 
the scalar $S$ vanishes. Unfortunately it would not mean that tensor
perturbations dominate because they contribute both into $\Phi$ and $\Psi$.

\section{Local and nonlocal description of polarization}

It is an interesting observational problem which quantity is easier to measure
in a noisy background, a differential local or an integrated global one. As
was stated in ref.~\cite{geom1} a measurement of an integrated quantity would 
be much more robust. And correspondingly the field variables $S$ and $P$ (or in
notations of papers~\cite{gw1}-\cite{gw3}, $E$ and $B$) were expressed as
integrals over all or a part of the sky. We think that the answer to the
question on the best observational strategy very much depends on the
properties of the noise. For example if the noise in polarization field of CMB
is created by point-like sources, chaotically distributed on the sky
with the mean separation larger than the resolution of the antenna, then
the measurement of local differential quantities, as e.g. direct
measurements of $S$ and $P$ given by eq. (\ref{sp}) seems easier. However
there may be sources of the noise that would be easier to suppress if one
measures a quantity which is averaged over whole (which is not possible) or
a part of the sky. To this end we re-derive the expressions for integrated
$S$ and $P$ (or $E$ and $B$) presented in ref.~\cite{geom2}.
The derivation is presented in great detail because of some disagreement
with ref.~\cite{geom2}. The results are very close but we show that
the window function may have a more general form even for the same choice
of normalization function $N(l^2)$, defined below.

Let us first define the Fourier transformed fields:
\be
\tilde {Q }(\vec l) = \int d^2 y \,e^{-i \vec l \,\vec y } \,Q (\vec y)
\label{ul}
\ee
and the similar one for $U$.
The Fourier transformed scalar and pseudo-scalar fields can be written as
\be
\tilde {S_N }(\vec l) =N(l^2) \int d^2 y \,e^{-i \vec l \,\vec y }
\left[Q(\vec y)\, \cos 2\phi_l  + U(\vec y)\, \sin 2\phi_l  \right]
\nonumber \\
\tilde {P_N }(\vec l) = N(l^2)\int d^2 y \,e^{-i \vec l \,\vec y }
\left[U(\vec y) \, \cos 2\phi_l -  Q(\vec y)\,\sin 2\phi_l \right]
\label{slpl}
\ee
where $\phi_l$ is the polar angle in the plane of Fourier coordinates
$\vec l$. The scalar function $N(l^2)$ is arbitrary. It preserves scalar
or pseudo-scalar property of $S$ and $P$. For the definition (\ref{sp}) one
has to choose $N(l^2) = l^2$. The definition used in ref.~\cite{geom2} is
$N(l^2) = 1$. This means that a non-locality is introduced in the coordinate
space by the inverse Laplace operator, $1/\partial^2$, that is by the
Green's function of the Laplacian.

Now we can make the inverse Fourier transform to obtain the functions
$S_N$ and $P_N$ in coordinate space:
\be
S_N(\vec x) = \int {d^2 l \over (2\pi)^2} N(l^2) \int d^2 y
e^{i\,{\vec l}\,(\vec{x}-\vec{y})}
\left[Q(\vec y)\, \cos 2\phi_l +U(\vec y)\,\sin 2\phi_l   \right]
\nonumber \\
P_N(\vec x) = \int {d^2 l \over (2\pi)^2} N(l^2) \int d^2 y
e^{i\,{\vec l}\,(\vec{x}-\vec{y})}
\left[-Q(\vec y)\, \sin 2\phi_l +U(\vec y)\,\cos2\phi_l   \right]
 \label{sxpx}
\ee
where $\phi_l$ is the angle between the vector $\vec l$ and some fixed
direction; it is convenient to choose the latter as the direction of vector
$\vec x$, so that $\phi_l \equiv \phi_{xl}$.

Integration over directions of vector $\vec l$ can be done explicitly. To
simplify the notations let us introduce
\be
\vec \rho = \vec x -\vec y
\label{rho}
\ee
and three angles $\phi_{l\rho}$, $\phi_{\rho x}$, and $\phi_{xl}$ between the
directions of the indicated vectors. Evidently
\be
\phi_{l\rho}+\phi_{\rho x}+ \phi_{xl} =0
\label{sumphi}
\ee
The angular integral is reduced to
\be
\int_0^{2\pi} d\phi_{l\rho} e^{il\rho \cos \phi_{l\rho} }
\left( A \cos 2\phi_{l\rho} + B \sin 2\phi_{l\rho} \right)
\label{intphilr}
\ee
where the coefficient functions $A$ and $B$ do not depend on $\phi_{l\rho}$.
The second term vanishes, while the first one gives
\be
\int _0^{2\pi} d\phi_{l\rho} e^{il\rho \cos \phi_{l\rho} }
\cos 2\phi_{l\rho} = -2\pi J_2 \left( l \rho \right)
\label{bessel}
\ee
where $J_2 (z)$ is the Bessel function (see e.g. \cite{gr}).

The integration over magnitude of $l$ depends upon the form of the function
$N(l^2)$ and the result is a function of the magnitude of the vector
$\vec \rho$:
\be
\int_0^\infty dl l  N(l^2) J_2 \left( l \rho \right) = F_N \left( \rho \right)
\label{frho}
\ee
For the particular case of $N(l^2) =1$ chosen in ref.~\cite{geom2} the
integral can be taken as follows. It is formally divergent so some
regularization procedure should be applied. This can be achieved by
introducing a small imaginary part  to $l$ to ensure convergence (in other
words, we have to shift the contour of integration to the upper $l$-half-plane). 
Using the relation~\cite{gr}:
\be
zJ_2(z) = J_1(z) -zJ_1' (z)
\label{zj2}
\ee
and integrating by parts, we obtain:
\be
F_1 \left( \rho \right) =
  {1\over \rho^2} \int_0^\infty dz z J_2(z)=
{1\over \rho^2} \left[ 2\int_0^\infty dz J_1(z) - zJ_1(z)|^\infty_0
\right] = {2\over \rho^2}
\label{f1}
\ee

Now taking all the contributions together we obtain:
\be
S_N(\vec x) =  {1\over 2 \pi} \int_0^\infty {d \rho \rho} F_N (\rho )
\int_0^{2\pi} d\phi \left[ Q\left(\vec x - \vec \rho \right) \cos 2 \phi +
 U\left(\vec x - \vec \rho \right) \sin 2 \phi \right]
\nonumber \\
P_N(\vec x) = {1\over 2 \pi} \int_0^\infty {d \rho   \rho} F_N (\rho )
\int_0^{2\pi} d\phi \left[- Q\left(\vec x - \vec \rho \right) \sin 2 \phi +
 U\left(\vec x - \vec \rho \right) \cos 2 \phi \right]
\label{spfin}
\ee
For the particular case of $ F_N (\rho ) = F_1 \left( \rho \right) =
{2/ \rho^2}$ considered in ref.~\cite{geom2} we obtain almost the same
result as the quoted paper with the only difference that we do not see
any reason to assume that the window function $F_1 (\rho) = 2/\rho^2$
should be taken zero at $\rho = 0$. Anyhow, this difference has zero 
measure and does not have any impact on the value of the 
integrals (\ref{spfin}). Hence it may be disregarded.
What, as we think, is more essential is the statement of ref.~\cite{geom2}
that in order to avoid difficult (or even impossible) integration of the data
over the whole sky one may use a modified window function:
\be
F_{ sz}(\rho) = - g(\rho) + {2\over \rho^2} \int_0^\rho d\rho' \rho' g(\rho')
\label{fsz}
\ee
with the function $g(\rho)$ subject to the condition:
\be
\int d\rho \rho g(\rho) =0
\label{intg}
\ee
where the last integral is taken over all the sky.

We believe that any window function can be used and no additional
conditions are necessary. To show that we calculate the functions
$S_N(\vec x) $ and $P_N(\vec x)$ for the particular case of scalar
perturbations when the Stokes matrix is given by 
expression~(\ref{apsi}). Calculation of derivatives in polar coordinates is
straightforward and after some algebra we obtain:
\be
S_N (\vec x) = {1\over 2\pi} \int_0^\infty d\rho \rho W (\rho)
\int_0^{2\pi} d\phi \left( \Psi_{\rho,\rho}\left( \vec x -\vec \rho \right)
-{\Psi_{\rho }\left( \vec x -\vec \rho \right) \over \rho}
\right)
\nonumber \\
P_N (\vec x) = {1\over 2\pi} \int_0^\infty d\rho \rho W (\rho)
\int_0^{2\pi} d\phi
\left( {2\Psi_{\rho,\phi}\left( \vec x -\vec \rho \right) \over \rho} -
{2\Psi_{ \phi}\left( \vec x -\vec \rho \right) \over \rho^2} \right)
\label{sppsi}
\ee
where sub-$\rho$ or sub-$\phi$ means differentiation with respect to the
corresponding variable and $W(\rho)$ is an arbitrary window function.

One can see from the second of these expressions that indeed $P$ vanishes
for any window function. Thus to prove the absence of tensor perturbations,
one should either observe vanishing of the local quantity $P(\vec x)$
given by eq.~(\ref{spqu}) or of the nonlocal one given by eq.~(\ref{sppsi})
with an arbitrary convenient window function $W(\rho)$. Which method would
be more efficient depends upon the properties of the noise.

\section{Singular points in polarization maps}

\subsection{Introductory remarks}

Polarization state of photons of CMB can be described by the direction
of maximum polarization and its magnitude; the former is parallel to the
eigen-vector of the Stokes matrix and the latter is equal to
$\sqrt{ Q^2 + U^2}$. Polarization maps simulated in different papers present
the corresponding vector field on two-dimensional plane. For the analysis
of the flux lines on this map it is very important to know the properties of
the singular points of this vector field. It has been done in
ref.~\cite{ddnn} (see also \cite{nn}). Another approach 
was taken in refs~\cite{geom1,geom2} where the
properties of the flux lines were analyzed in terms of basis functions of
tensor spherical harmonics (see figs. 1 in refs.~\cite{geom1,geom2}).
However the flux lines of these basic functions are quite different
from the behavior of the flux lines of the polarization vector. Of course
the analysis in terms of tensor harmonics and the behavior of Stokes
parameters $Q$ and $U$ can be used for description of polarization maps but
the analysis in terms of eigenvectors of the Stokes matrix permits to make
a more direct description of the properties of the polarization field.
Possible types of different singular points as well as their statistical
distributions may bring a new piece of information about properties of CMB.
Of course a measurement of CMB polarization near the point where it
vanishes is a very difficult observational problem. However it is not
necessary to go exactly to the point where $Q^2+U^2 =0$. The type of
the singularity can be determined by the pattern created by the flux 
lines in the region where polarization is non-vanishing (see an example
of simulated polarization map below in Fig. 4).
 
The analysis of singular points of the polarization vector field was performed
in ref.~\cite{ddnn} and \cite{nn}. It was found in \cite{ddnn} that their 
types do not fit the well known classification of singular points of vector 
fields in the standard theory of dynamical systems. Due to non-analytic 
behavior of the eigen-vectors near the the zero points, $Q^2+U^2 =0$, the 
separatrices end at the singularity, while in the usual case they smoothly 
continue through these points. These unusual behavior, found in our 
paper~\cite{ddnn}, is well observed in the polarization maps simulated 
in refs.~\cite{geom2} and in the maps of our paper below(Figs. 4,5). 
In this and the next section we present a further development of the
analysis of our previous shorter paper~\cite{ddnn}.

\subsection{Basic equations}

The eigenvectors of polarization matrix (\ref{aqu}) are:
\be
\vec n\,^+ \sim \{ U, \lambda - Q \} \nonumber \\
 \vec n\,^- \sim \{ -U, \lambda + Q \}
\label{vecn}
\ee
where $\lambda =\sqrt{ Q^2 + U^2}$ is the magnitude of the eigenvalue and 
the vectors $\vec n\,^{\pm}$ correspond respectively to the positive and 
negative eigenvalues, $\pm \lambda$. The vector $\vec n\,^+$ is parallel
to the direction of the maximum polarization, while $\vec n\,^-$ goes
along the direction of the minimal polarization. This is evident in the
basis of eigenvectors where the polarization matrix is diagonal,
${\mathbf{a}} = {\rm diag} \{ \lambda, -\lambda \}$. The total intensity of
light polarized along $\vec n\,^{\pm}$ is given by $I_\pm = I_0 \pm \lambda$.
Thus the intensity along $\vec n\,^+$ is bigger.

For definiteness we will consider the field of directions of the vector
$\vec n\,^+$ and the singular points in this field. The problem of singular 
points of a vector field $\vec V$ is investigated for the case when the
direction of this two-dimensional vector field with the components
$ \left[ x(t), y(t) \right]$ is governed by the equation
\be
{dy\over dx} = {F_1(x,y) \over F_2(x,y) }
\label{fj}
\ee
Singularities may appear if simultaneously both functions 
$F_{1,2}(x,y)$ vanish. In this case the conditions of uniqueness of the
solution of the differential equation is not fulfilled and more than one
integral curve may pass through the same point.
The standard theory is developed for the case when the
functions $F_{1,2}$ are analytic near these zeroes, and their
first order Taylor expansion has the form:
\be
F_j = a_j (x-x_0) + b_j (y -y_0)
\label{ftayl}
\ee 
The following three singular points are possible in this case: knots,
saddles, and foci (see e.g.~\cite{bs}). The separatrices of the solutions 
are two intersecting
lines, which are simply straight lines in the linear approximation.  

However in the case of polarization vector field the basic equation has the 
form:
\be 
{dy\over dx} ={ n^{+}_y \over n^+_x} = { \lambda -Q \over U}
\label{dydx}
\ee
The singular points may appear, as above, if both numerator and denominator
vanish. It is equivalent to the condition $Q=U=0$. An essential difference to 
the standard case is that now the numerator is not analytic near zero. This
fact results in a quite different behavior of the integral curves near these
points. The standard theory is not applicable to this case and below we will
investigate the structure of solutions in the vicinity
of these points directly. We assume that
the functions $Q$ and $U$ are analytic near the points where $Q=U=0$, so that
they can be  expanded as: 
\be
Q \approx q_1 x + q_2 y \nonumber  \\
U \approx u_1 x + u_2 y
\label{quexp}
\ee
For the sake of brevity we assume that $Q$ and $U$ vanish at $x=y=0$. 

\subsection{Types of singular points}

It is convenient to introduce the new coordinates:
\be 
\xi = q_1 x + q_2 y \nonumber  \\
\eta = u_1 x + u_2 y
\label{xieta}
\ee
Since this coordinate transformation corresponds to a rotation and rescaling
of the coordinates, the forms of singular points would remain the same. Now 
we introduce polar coordinates on the plane $(\xi, \, \eta) $:
\be
\xi = r \cos \phi, \,\,\, \eta = r \sin \phi
\label{rphi}
\ee
In these coordinates the equation (\ref{dydx}) is rewritten as
\be
{d\ln r \over d\phi} = {N  \over D} \equiv
{ q_2 t^3 +(q_1 -2u_2) t^2 - (q_2 +2u_1) t -q_1 \over
u_2 t^3 +(u_1 + 2q_2) t^2 + (2q_1 - u_2) t -u_1 }
\label{dlnr}
\ee
where $t= \tan (\phi/2)$.

In the general case the denominator $D$ has three roots $t_j$, $j=1,2,3$. 
Without loss of generality we may assume that $u_2 =1$. Then these roots 
satisfy the conditions:
\be
t_1 t_2 t_3 = u_1 ,\nonumber \\
t_1 t_2  +t_2 t_3  + t_3 t_1 = 2q_1 - 1 , \nonumber \\
t_1 + t_2 + t_3 =-(u_1 + 2q_2)
\label{tj}
\ee
The integration of equation (\ref{dlnr}) becomes straightforward if we 
expand the r.h.s. in elementary fractions:
\be
{d\ln r \over d\phi} = q_1 + \sum_j^3 {B_j \over t-t_j}
\label{gam}
\ee
where, as one can easily see,  
$B_j = N(t_j) /(t_j-t_k)(t_j-t_l)$, none of $j,k,l$ are equal to any of the
others. It is straightforward to verify that
\be
B_1 = -{ (1+t_2 t_3)(1+t_1^2)^2 \over 2 (t_1-t_2)(t_1-t_3) }
\label{b1}
\ee
Remaining parameters $B_2$ and $B_3$ are obtained by cyclic permutations.

Since $d\ln r / d\phi =  (d\ln r / dt) (1+t^2)/2$ the equation can be 
finally rewritten as
\be
{d\ln r \over  dt} = {2\over 1+t^2} \left( q_1 + 
\sum_j^3 {B_j \over t-t_j} \right)
\label{eqnfin}
\ee
and the integration becomes straightforward. The corresponding solution is: 
\be
r = r_0 \left( 1+t^2\right) \prod_j^3 \left(t-t_j \right)^{2\nu_j}
\label{sol}
\ee
where $r_0$ is an arbitrary constant and the powers $\nu_j$ are
\be
\nu_j = {B_j \over 1+t_j^2 }
\label{nuj}
\ee
with the constants $B_j$ given by eq. (\ref{b1}).
It can be checked that $\nu_j$ satisfy the following conditions:
\be
\sum_j^3 \nu_j = -1 , 
\label{sumnu}
\ee
\be
\sum_j^3 \nu_j t_j =-{1\over 2}\left( \sum_j^3  t_j + \prod_j^3 t_j\right)=
q_1 ,
\label{sumnut}
\ee
\be
\prod_j^3 \nu_j = { (1+t_1^2)(1+t_2^2)(1+t_3^2) \over
8(t_1-t_2)^2 (t_2-t_3)^2 (t_3-t_1)^2 }
(1+t_1t_2)(1+t_2t_3)(1+t_3t_1)
\label{prodnu}
\ee
The last three factors in equation (\ref{prodnu}) are proportional to 
the determinant $d=q_1 u_2 - q_2 u_1$:
\be
(1+t_1t_2)(1+t_2t_3)(1+t_3t_1)= 2(q_1 u_2 - u_1 q_2)/u_2^2 \equiv 2d/u_2^2
\label{det}
\ee
If all the roots $t_j$ are real, then the sign of the product $\prod_j^3 \nu_j$
is the same as the sign of the determinant $d$. If however one of the roots,
e.g. $t_1$, is real and the other two are complex conjugate, the sign of the
determinant and of the product (\ref{prodnu}) are opposite.

Now we can make the classification of the singular points. Let us first
consider the case when all the roots $t_j$ are real. The behavior of the
solution is determined by the signs of the powers $\nu_j$. Due to  
equation (\ref{sumnu}) at least one of the powers $\nu_j$ must be
negative. To see what other 
signs are possible let us assume (without loss of generality) that
\be
t_1>t_2>t_3
\label{ineq}
\ee
In this case the following sign relations are valid:
\be
{\rm sign} [\nu_1] = {\rm sign} \left[ - (1+t_2t_3) \right], \nonumber \\
{\rm sign} [\nu_2 ]= {\rm sign} \left[ (1+t_1t_3) \right], \nonumber \\
{\rm sign} [\nu_3 ]= {\rm sign} \left[ - (1+t_1t_2) \right]
\label{signs}
\ee
If e.g. $t_3 >0$, the following signs of $\nu_j$ are realized $(-,+,-)$.
If $t_3 <0$ but  $t_2 >0$, then $\nu_3 <0$ and one or both $\nu_1$ and $\nu_2$
are negative. They cannot both be positive because if $ (1+t_1t_3) >0$, then
$(1+t_2t_3) >0$ too and $\nu_1 <0$. Analogously, in the case $t_1 >0$ and
$t_2 <0$ the set of signs $(-,+,+)$ for any sequence of $\nu_j$ is impossible.
In the case when all $t_j$ are negative, the sign pattern is $(-,+,-)$. Thus
only two sign combinations for $\nu_j$ are possible: $(-,-,-)$ and
$(-,-,+)$. The first one is realized when $d<0$ in accordance with 
expressions (\ref{prodnu}) and (\ref{det}). If the determinant is positive, 
then the signs of $\nu_j$ are $(-,-,+)$. 

In the case when $d<0$ the solution
does not pass through zero in the vicinity of the singular point. 
Its behavior is similar to the usual saddle with the
only difference that there are three and not four, as in the usual case,
linear asymptotes/separatrices (see Fig. 1a). We will also call it a "saddle".
The fact that in our case separatrices are not continued through the singular
point, in contrast to the usual singularities is related to the non-analytic
behavior of the equation (\ref{dydx}) due to the square root singularity.

If $d>0$, then the sign pattern is $(-,-,+)$ and the solution vanishes
along one of the directions and tends to infinity along the other two. The 
form of the solution is quite different from the standard ones. 
The field line cannot be continued along $\phi =\phi_1$ into
$\phi =\phi_1 +\pi$ as can be done in the usual case. We will
call this type of singularity a "beak" (see Fig. 1b).

If only one of the roots $t_j$ is real and the other two are complex 
conjugate, the solution has the form:
\be
{r\over r_0} =(t^2 +1) \mid t - t_2 \mid ^{4{\rm Re}\, \nu_2}
\exp \left(4 \beta\, {\rm Im} \,\nu_2   \right)
\left( t-t_1 \right)^{2\nu_1}
\label{r1}
\ee
where $\beta = \tan^{-1} [ {\rm Im} t_2 /( t-{\rm Re} t_2 )]$.
The real root $\nu_1$ is negative, as is seen from eq. (\ref{prodnu}) and
thus $r$ does not vanish in vicinity of such singular point. The flux lines
of the polarization field for this case are presented in Fig. 1c. This
type of singularity can be called a "comet". This case is realized
when determinant $d$ is positive.

\begin{figure}
\centering
\epsfxsize=15cm
\epsfbox{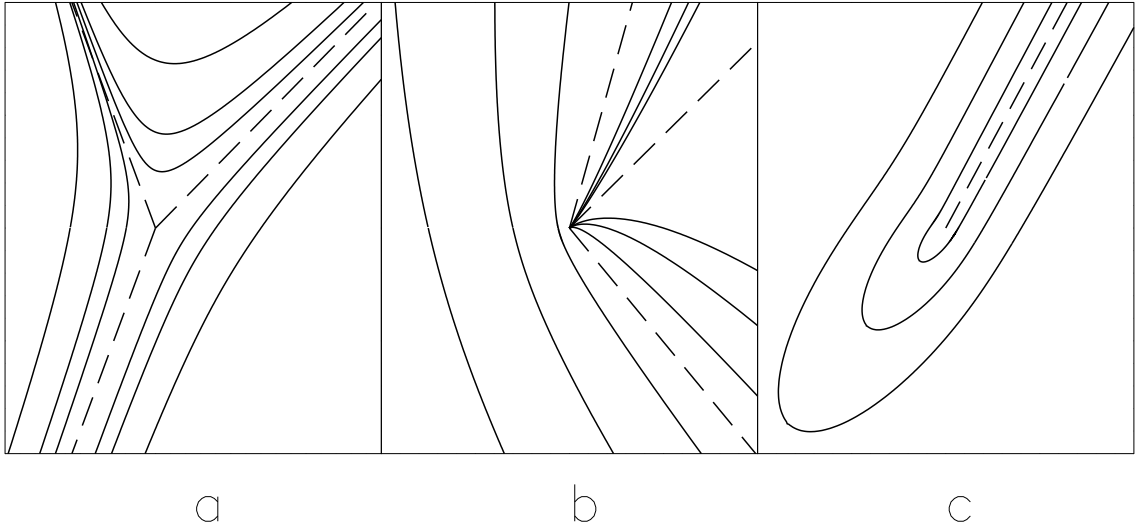}
\vspace{0.75cm}
\caption{Flux lines for three different 
types of singular points: 
(a) saddle, (b) beak, and (c) comet. Dashed lines show 
peculiar solutions (separatrices).} 
\end{figure}

\subsection{Probabilities of various types of singularities}

The relative weights of different singular points was calculated in the
following way. It is evident that the probability of "saddles" is 
50\% because saddles appears if and only if $d<0$. The probability of 
comets and beaks was found numerically from the conditions that $d>0$ and
there is a single real root of the equation $D=0$ (for comets)
or there are three real roots (for beaks), 
where $D$ is the denominator in the expression (\ref{dlnr}). The 
probability of appearance of saddles, beaks and comets for random choice
of $q_1$, $q_2$, $u_1$, and $u_2$, is correspondingly $W_s=0.500$,
$W_b\approx 0.116$, $W_c\approx 0.384$.

One can also estimate the number density of the singular points in the
following way (see e.g. \cite{be,nn}). All singular points correspond to
the case when both $Q=0$ and $U=0$. The  number density of these points
is proportional to
\be
dQdU =  |d| dx dy
\label{dqdu}
\ee
and thus the density is given by the average value of the determinate,
$d= q_1 u_2 - q_2 u_1$. 
It can be shown that saddles make 50\% of all singular points
$\langle n_s\rangle =0.5\langle n \rangle$, where $n$ is the number
 density of all singular points.
Calculations of the number density of beaks and comets are more complicated 
and should be done numerically.
According to our estimates the surface densities for beaks and comets
are correspondingly $\langle n_b \rangle \approx 0.052\langle n \rangle$
and $\langle n_c\rangle \approx 0.448 \langle n \rangle$.
Deviations from these and above found numbers for $W_{s,b,c}$ may signal 
deviations from Gaussian nature of perturbations.

\section{Comparison of two classification methods}

As we have mentioned above the first classification of singular points
in CMB polarization map was performed in ref.~\cite{nn}, where the 
equation 
\be
{dy\over dx} = {Q\over U}
\label{dydxQU}
\ee
was used to describe the behavior of the flux lines of the "vector"
$\vec V = [U,Q]$. However under coordinate transformations $U$ 
and $Q$ are not transformed as components of a vector but as 
components of a second rank tensor in accordance with eq.~(\ref{q'u'}).
Because of that the maps of the flux lines of the "vector"
$\vec V$ would not be invariant with respect to rotation of the coordinate
system. In a fixed reference frame there could be three possible types
of singular points in accordance with the standard classification~\cite{bs}:
knots, foci, and saddles. This is so because both functions $Q$ and $U$
generically should be analytic near the point where both of them vanish.
The types of singularities in these classification scheme depend in particular
upon the sign of the determinant $d= q_1 u_2 -q_2 u_1$ introduced above.
If $d>0$, then the singular points are saddles, while for $d<0$ there may 
be both foci and knots. 

It can be shown that $d$ is invariant under coordinate rotation. Thus 
saddles retain their identity in different coordinate frame. On the other
hand, foci and knots may transform into each other under rotation. So the
topology of the map of flux lines of the "vector" $\vec V$ would look
differently in different coordinate systems, though the positions of
the singular points evidently remains the same. Moreover the flux lines
of another possible "vector", $ \vec W = [Q,U]$ are in a sense 
complementary
to those of $\vec V$. The relevant determinant changes sign and thus saddles
of $\vec V$ correspond to foci and knots of $\vec W $ and vice versa.

In contrast to classification based on $\vec V$ or $\vec W $, the 
description of the polarization map in terms of eigenvectors of the
Stokes matrix, considered in the previous sections and in 
ref.~\cite{ddnn}, is invariant with respect to coordinate transformations,
so that types of the singular points do not change under rotation.
As we have argued in the previous section, positive $d$ gives rise to either
to beaks or comets, while negative $d$ could produce only saddles. Thus
a saddle in the flux lines of $\vec n\,^{(+)}$ corresponds to a knot or 
a focus on $\vec V$-map and to a saddle on $\vec W$-map. A comet or a beak
on $\vec n\,^{(+)}$-map both correspond to a saddle on $\vec V$-map or to
either a knot or a focus on $\vec W$-map. The latter relations may be
different in different coordinate frames because, as we mentioned above, 
knots and foci transforms into each other under rotation, while beaks and
comets remain the same (this is also true for saddles on $\vec n\,^{(+)}$-map).
These statements are illustrated on Figs. 1-3 where examples of
the maps for $\vec n\,^{(+)}$ and for $\vec V$ are presented.

\begin{figure}
\centering
\epsfxsize=11.0 cm
\hskip 1.5cm
\epsfbox{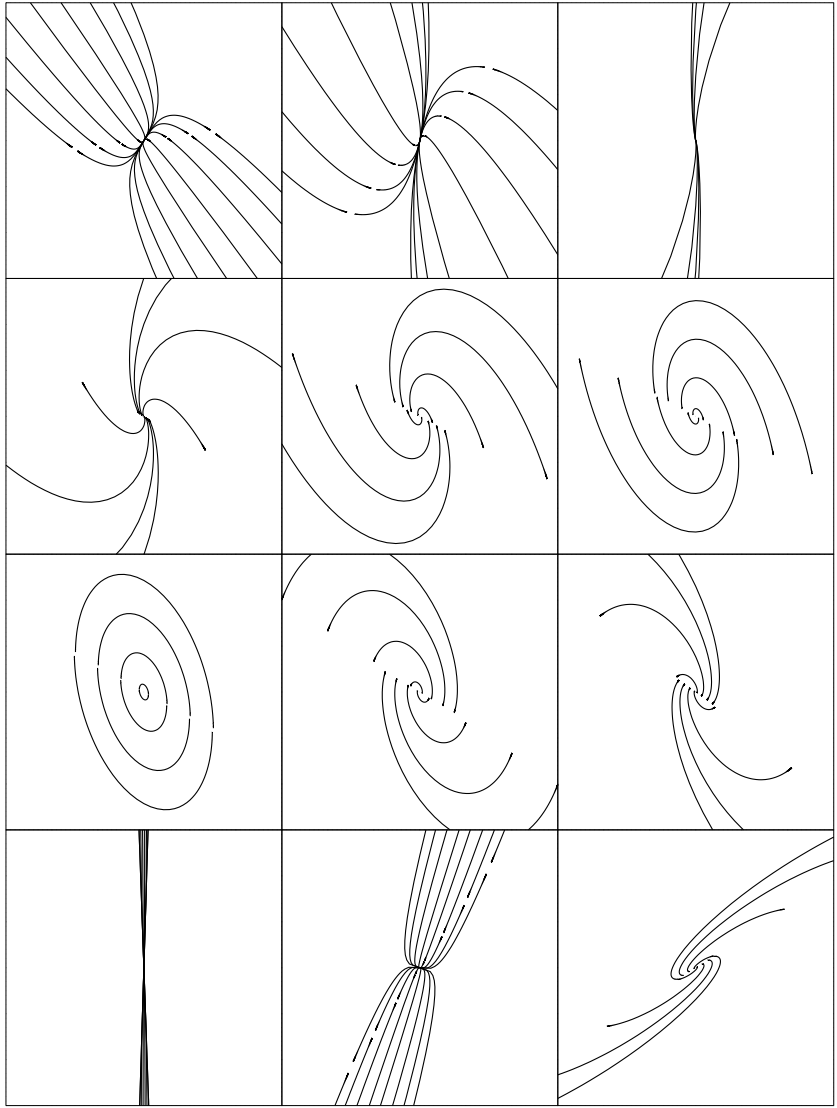}
\vspace{0.25cm}
\caption{Transformation of the flux lines of "vector" $\vec V$ due to 
rotation of coordinates near the saddle type singularity of $\vec n^{(+)}$,
plotted in Fig. 1(a). The set of maps from left to right and from the
top to the bottom corresponds to rotations with respect to the first one
by the angles: 
$\phi= 5^\circ, 10^\circ, 15^\circ, 20^\circ, 22.5^\circ, 25.6^\circ, 30^\circ, 
34^\circ, 49^\circ, 57^\circ,\,\, {\rm and}\,\, 70^\circ$. 
} 
\end{figure}

\begin{figure}
\centering
\epsfxsize=14 cm
\epsfbox{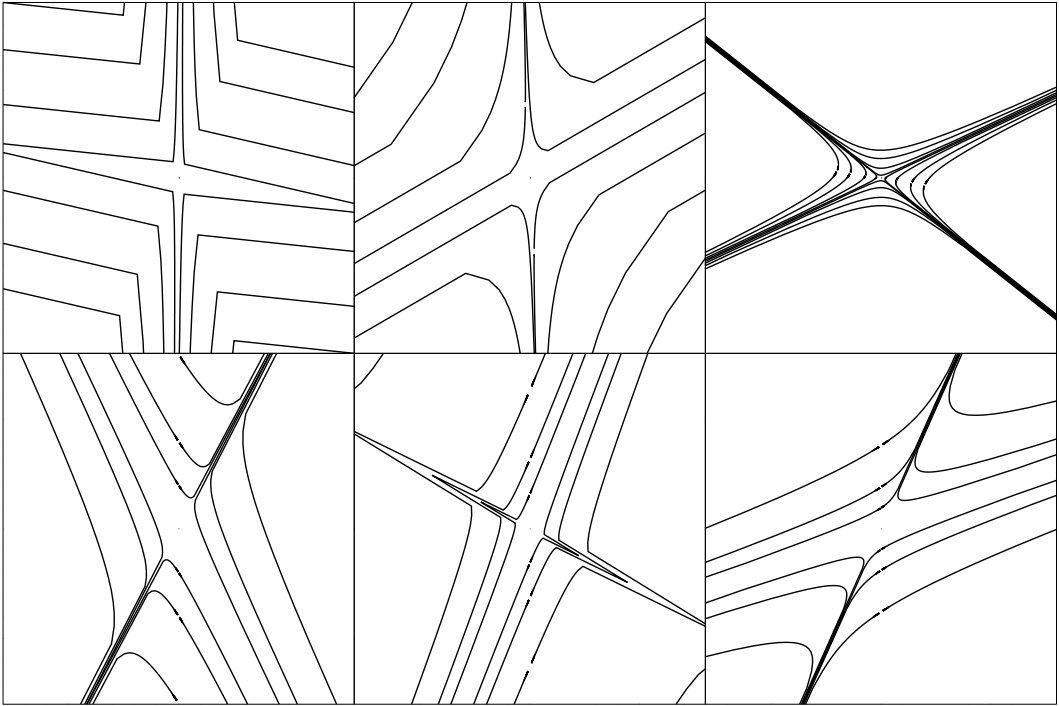}
\vspace{0.5cm}
\caption{Transformation of the flux lines of the "vector" $V$ near the 
beak (upper panel) and comet (lower panel) type singularities of 
$\vec n^{(+)}$, plotted in Fig. 1(b) and 1(c) respectively.
The maps from left to right correspond to rotation by
$\phi= 45^\circ\,\,{\rm and}\,\,  90^\circ$ with respect to the first one.}
\end{figure}

The simulated 500*500 pixels, $5^\circ \times 5^\circ$ degree map of the 
CMB polarization field for the Standard CDM model with the HWFM
resolution $0.3^\circ$ is presented in Fig.4.
Each vector represents orientation of the linear polarization with the 
angle $\frac{1}{2} {\tan}^{-1}(U/Q)$ counterclockwise
to the positive direction of the X axes. Length of each
vector is proportional to $\sqrt{Q^2+U^2}$.
For visual clarity we use only 50*50 pixels.
Solid lines represent the behavior of the field in the
vicinity of non-polarized points. In this picture we
have 4 saddles, 3 comets and 1 beak. For comparison of two approaches
described in this section the same random Gaussian realization as for
Fig. 4 but for the pseudo vector $\vec V$ is
presented in Fig. 5. As we discussed above the types of the singularities 
in this case are different. There are 4 saddles which correspond to  
comets or beaks in the polarization map of Fig. 4 as well as 
2 knots and 2 foci. The latter can only be the saddles in the
polarization pattern of Fig. 4.

\begin{figure}
\centering
\epsfxsize=10 cm
\epsfbox{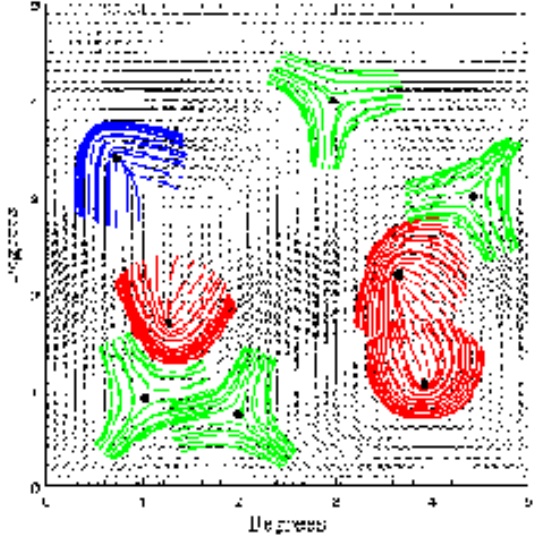}
\vspace{0.5cm}
\caption{Simulated map of CMB polarization vector field $\vec n^{(+)}$
(see Sec. 5 for details). Solid lines show the flux line behavior near
singular points where polarization vanishes.
}
\end{figure}

\begin{figure}
\centering
\epsfxsize=14 cm
\epsfbox{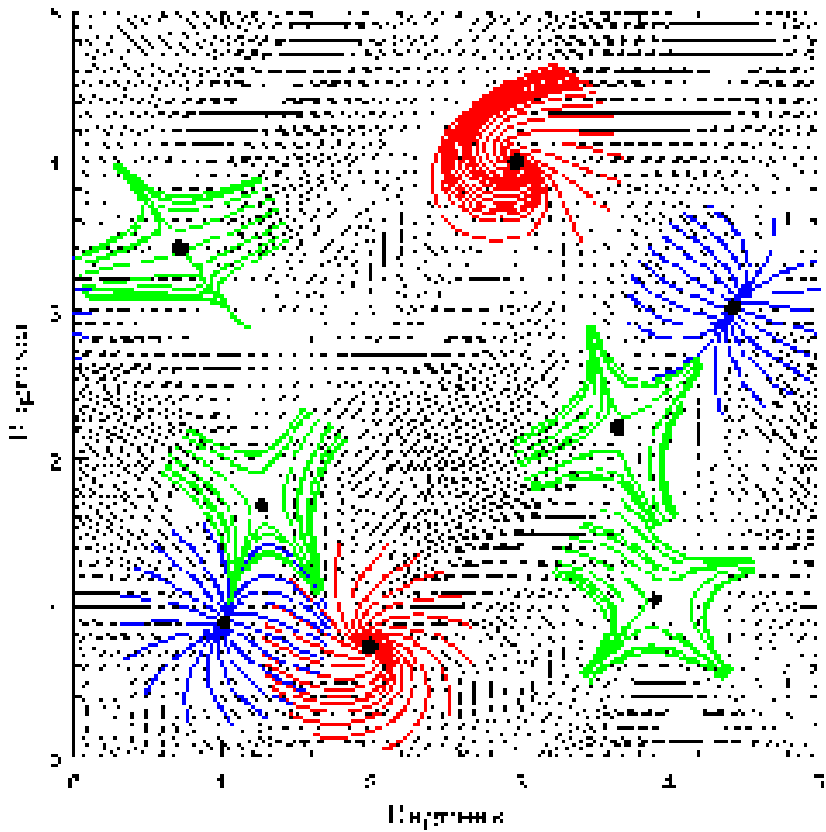}
\vspace{0.3cm}
\caption{Simulated map of the flux lines of "vector" $\vec V$ (see sec. 5 for 
details). Solid lines show the flux line behaviour near
singular points where polarization vanishes.
} 
\end{figure}
 
\section{Minkowski Functionals.}
 
In this section we will consider an application of Minkowski Functionals (MF)
for description of statistical properties of CMB polarization. For two 
dimensional maps the following three MF's are of interest:
area ($A$), length ($L$), and genus ($G$). 
As usually, we will consider the magnitude of polarization, 
$\sqrt{Q^2+U^2}$ as a two-dimensional surface in a three-dimensional 
space. If we cut this surface at different levels $P_t$, then 
the area of the map will be divided into two parts: the area where 
polarization is above the threshold $P_t$ and the area where $P<P_t$. 
Three functions characterize this division:
\begin{enumerate}
\item{} 
"Area", $A$, is the fraction of the area of the map where $P>P_t$;
\item{}
"Length", $L$, is the surface density of length of the boundary 
between the fractions with $P>P_t$ and $P<P_t$;
\item{}
"Genus", $G$, is the number density of isolated highly polarized 
regions minus the number density of isolated weakly polarized
regions (equivalent to the Euler characteristic).
\end{enumerate}
All three Minkowski Functionals, $A$, $L$, and $G$, are evidently 
functions of $P_t$.

Minkowski Functionals (MF) have some mathematical properties that 
make them especially convenient for our task. They are translationally 
and rotationally invariant, additive, and have simple and intuitive 
geometrical meaning. Moreover, it was shown~\cite{min,had} that 
morphological properties of any pattern in $D$-dimensional 
space, which possess the above mentioned characteristics, can be fully 
described by $D+1$ Minkowski Functionals. The CMB polarization, 
that we consider, is a two-dimensional random field, so the introduced 
above three MF's would fully characterize these properties. 
MF's are additive with respect to isolated parts of the sky that
makes them potentially very powerful for patchy coverage.  
These properties make MF's a convenient tool for checking the 
polarization patterns for a presence of a non-Gaussian component. 

For statistical analysis of CMB polarization field it is convenient 
to express the Stokes parameters $Q$ and $U$ in terms of scalar and 
pseudo-scalar potentials $\Psi$ and $\Phi$ as given by eq. 
(\ref{ageneral}):
\begin{equation}
Q=\Phi_{11}-\Phi_{22}-2\Psi_{12}, 
\quad U=2\Phi_{12}+\Psi_{22}-\Psi_{22}
\label{quphipsi}
\end{equation}
where $\Phi_{ij}$, $\Psi_{ij}$ are the second derivatives of $\Phi$ 
and $\Psi$ in a Cartesian coordinate system in the small (flat) part
of the sky, as is described above. The potentials $\Phi$ and $\Psi$ 
can be considered as two independent two-dimensional Gaussian fields. 
For description of their statistical properties 
it is convenient to use their Fourier representation:
\be 
\Phi(\vec {x})={1\over 2\pi}\int C_{\Phi}(k)e^{-i\vec {k}\vec {x}}
d\vec {k} \nonumber \\
\Psi(\vec {x})={1\over 2\pi}\int C_{\Psi}(k)e^{-i\vec{k}\vec{x}}d\vec{k}
\ee

Statistical properties of the fields $Q$, $U$ and their 
derivatives are characterized by dispersions:
\be
\sigma^2_Q=\frac{1}{2\pi}\left[\int_0^{2\pi}d\theta\cos^2(2\theta)
\int_0^\infty k^5|C_\Phi(k)|^2dk+\int_0^{2\pi}d\theta\sin^2(2\theta)
\int_0^\infty k^5|C_\Psi^2(k)|^2dk\right], \nonumber \\
\sigma^2_U=\frac{1}{2\pi}\left[\int_0^{2\pi}d\theta\cos^2(2\theta)
\int_0^\infty k^5|C_\Psi(k)|^2dk+\int_0^{2\pi}d\theta\sin^2(2\theta)
\int_0^\infty k^5|C_\Phi(k)|^2dk\right], \nonumber \\
\quad \sigma^2_Q=\sigma^2_U={1\over 2}[\sigma_{0,\Psi}^2+ 
\sigma_{0,\Phi}^2], \hskip 5cm 
\\
\sigma^2_{0,\Phi}=\int_0^\infty k^5 |C_\Phi(k)|^2dk, \quad 
\sigma^2_{0,\Psi}=\int_0^\infty k^5 |C_\Psi(k)|^2dk,\nonumber \hskip 2cm\\
\sigma^2_{1,\Phi}=\int_0^\infty k^7 |C_\Phi(k)|^2dk, \quad 
\sigma^2_{1,\Psi}=\int_0^\infty k^7 |C_\Psi(k)|^2dk,\nonumber\hskip 2cm \\
\sigma^2_{2,\Phi}=\int_0^\infty k^9 |C_\Phi(k)|^2dk, \quad 
\sigma^2_{2,\Psi}=\int_0^\infty k^9 |C_\Psi(k)|^2dk.\hskip 2cm \nonumber 
\label{sigman}
\ee
Expressions for the correlators of first and second derivatives, namely, 
$Q_i, Q_ij, U_i, U_{ij}$, through $\sigma^2_{1,\Phi}, \sigma^2_{1,\Psi}, 
\sigma^2_{2,\Phi}\,\, {\rm and}\,\, \sigma^2_{2,\Psi}$, are presented below.

Two important characteristics of the polarization are the autocorrelation 
radius, $r_c$, and the coefficient of cross correlation between functions 
$Q$ and $U$ and their second derivatives, $\gamma$:
\be
r_c^2=\frac{\sigma_{0,\Psi}^{ 2}+ \sigma_{0,\Phi}^{ 2}}
{\sigma_{1,\Psi}^{ 2}+ \sigma_{1,\Phi}^{ 2} },
\label{rc}
\ee
\be
\gamma=\frac{\sigma_{1,\Psi}^{ 2}+ \sigma_{1,\Phi}^{ 2}}
{\left(\sigma_{0,\Psi}^{ 2}+ \sigma_{0,\Phi}^{ 2} \right)^{1/2}
\left(\sigma_{2,\Psi}^{ 2}+ \sigma_{2,\Phi}^{2} \right)^{1/2}}.
\label{gamma}
\ee 

It is convenient to use instead of $Q$ and $U$ and their derivatives
dimensionless variables, normalized to their dispersions:  
\be
q= {Q \over \sqrt{\left( \sigma_{0,\Psi}^{ 2}+ \sigma_{0,\Phi}^{ 2}
\right)/2}},\quad 
u= {U \over \sqrt{\left( \sigma_{0,\Psi}^{ 2}+ \sigma_{0,\Phi}^{ 2}
\right)/2}},\nonumber \\
q_i = {Q_i \over \sqrt{\left(\sigma_{1,\Psi}^{2}+ \sigma_{1,\Phi}^{ 2}
\right)/2}},\quad 
u_i = {U_i \over \sqrt{\left(\sigma_{1,\Psi}^{2}+ \sigma_{1,\Phi}^{ 2}
\right)/2}},\\
q_{ij}= {Q_{ij} \over \sqrt{\left(\sigma_{2,\Psi}^{2}+ 
\sigma_{2,\Phi}^{ 2}\right)/2}},\quad 
u_{ij}= {U_{ij} \over \sqrt{\left(\sigma_{2,\Psi}^{2}+ 
\sigma_{2,\Phi}^{ 2}\right)/2}}.\nonumber 
\label{dimless}
\ee
Below we calculate Minkowski Functionals using these 
dimensionless variables. It is noteworthy that while 
$Q_i \equiv \partial Q/\partial x^i $, its dimensionless analogue, 
$q_i$, is not equal to $\partial q/\partial x^i$ 
but differs from it by the constant factor $r_c$. Thus, for example, 
\be
q_i = r_c \partial_i q,\quad q_{ij} = r_c^2 \partial_i\partial_j q
\label{qidq}
\ee 

\subsection{Area}

The simplest Minkowski Functional, that is the fraction of area of the 
map with $p>p_t$, can be calculated using the joint probability 
distribution function (PDF) $B(q,u)$ for values of $q$ and $u$ only. Since 
$q$ and $u$ are Gaussian fields, this function can be written using 
correlators for $q$ and $u$ and their dispersions:
\be
<q>=<u>=<qu>=0,\hspace{0.5cm} <q^2>=<u^2>=1
\label{cordis}
\ee
and we have 
\be 
B(q,u)dqdu=\frac{1}{2\pi}\exp\left(-\frac{q^2}{2}-
\frac{u^2}{2}\right)dqdu\, ,
\label{bqu}
\ee
The PDF of the amplitude 
$p=\sqrt {q^2+u^2 }$ is:
\be 
B(p)dp=p \exp\left(-\frac{p^2}{2}\right)dp
\label{bpdp}
\ee
and the relative area of the map where $p>p_t$ is:
\be 
A(p_t)=\int\limits^{\infty}_{p_t}B(p)dp = \exp \left(-\frac{p_t^2}{2}\right)
\label{area}
\ee

\subsection{Length}

The condition 
\be
p(x,y) = p_t
\label{yyxp}
\ee 
describes a curve in two-dimensional plane. The mean density of the length 
of this curve can be found from the expressions 
\be
dp(x,y) = (\nabla p ~d\vec{l}) = \sqrt{p_x^2+p_y^2}\,\,{dl\over r_c},\\
L(p_t) = {1\over r_c}\langle \sqrt{p^2_x+p^2_y}\,\rangle \,B(p_t)
\label{lpt}
\ee
where $\langle \sqrt{p^2_x+p^2_y}\,\rangle$ must be found for a 
given $p = p_t$. 

To do this we should use the PDF $B(p_t,p_x,p_y)$. For Gaussian $q$ and $u$
this function would also have a Gaussian form and to find it
we need correlators of  $q$ and $u$ and their first
derivatives:
 \be 
<qq_i>=<uu_i>=<qu_i>=<uq_i>=0,\nonumber  \\
<q_i>=<u_i>=<q_iu_j>=0, \\
<q_iq_j>=<u_iu_j>=\frac{1}{2}\delta_{ij}\nonumber
\label{qqi}
\ee
where $\delta_{ij}$ is the Kronecker symbol.
Using  this correlators we obtain
\be 
B(p,p_x,p_y)dpdp_xdp_y=\frac{1}{\pi}p\exp\left(-p^2/2-p^2_x-p^2_y
\right) dpdp_xdp_y, \\
\langle \sqrt{p^2_x+p^2_y}\,\rangle = {\sqrt{\pi}\over 2}
\label{bppxpy}
\ee
and, finally, 
\be 
L(p_t)=\frac{p_t\sqrt{\pi}}{2 r_c} \exp \left(-\frac{p_t^2}{2} \right)
\label{lptf}
\ee 

\subsection{Genus}

As is well known, Genus, $G(p_t)$ is expressed through the conditional 
mean value of determinant $det(p_{ij})=p_{11}p_{22}-p_{12}^2$ under 
conditions $p\geq p_t,\,\, p_i = 0$:  
\be 
G(p_t)={1\over r_c^2}\int\limits^{\infty}_{p_t}dp
\int\limits^{\infty}_{-\infty} \prod dp_{ij}\det (p_{ij}) B(p,p_i=0,p_{ij})
= {2\over \pi \gamma^2 r_c^2}\int\limits^{\infty}_{p_t}dpB(p)
\langle \Delta(p)\,\rangle
\label{gpt}
\ee
where $B(p,p_i=0,p_{ij})$ is the conditional PDF of its arguments and 
$\langle \Delta(p)\,\rangle $ is the {\it conditional} mean of  
$det(p_{ij})$. To find 
$ \langle \Delta(p)\,\rangle $ we should know the {\it conditional} mean 
value and dispersions of $p_{ij}$ which can be expressed through the 
{\it conditionless} mean value and non-vanishing correlators of $q_{ij}$ 
and $u_{ij}$, namely: 
\be
<qq_{ii}> = <uu_{ii}>=-\frac{\gamma}{2}\delta_{ij}, \nonumber \\
<q_{ii}^2> = {1\over 16}~ { 7 +5 \rho^2 \over 1 + \rho^2 }, \nonumber \\
<u_{ii}^2> ={1\over 16}~ { 5 +7 \rho^2 \over 1 + \rho^2 }, \nonumber \\
<q_{11}q_{22}>=<q_{12}^2> =
{1\over 16} ~{ 1 +3 \rho^2 \over 1 + \rho^2 }, 		 \\ 
<u_{11}u_{22}>=<u_{12}^2>= 
{1\over 16}~ { 3 +\rho^2 \over 1 + \rho^2 }, \nonumber \\ 
<q_{11}u_{12}>=<u_{11}q_{12}>=
{1\over 16} ~{ 1 -\rho^2 \over 1 + \rho^2 }, \nonumber \\ 
<q_{22}u_{12}>=<u_{22}q_{12}>= 
{1\over 16} ~{ -1 +\rho^2 \over 1 + \rho^2 }, \nonumber 
\label{qqii}
\ee
where $\rho^2 = \sigma_{2,\Psi}^2 / \sigma_{2,\Phi}^2 $. These correlators
are linked to correlators $p_{ij}$ through the expressions
\be 
p_{ij}={q\over p}q_{ij}+{u\over p}u_{ij}+
w_{ij},\quad w_{ij} = {1\over p}(q_{i}q_{j}+u_{i}u_{j}-p_{i}p_{j})
\label{ppij}
\ee
The required determinant $\langle \Delta(p)\,\rangle$ can be written as  
\be 
\langle \Delta(p)\,\rangle  = {q^2\over p^2}
\langle q_{11}q_{22}-q_{12}^2 \,\rangle + 
{u^2\over p^2}\langle u_{11}u_{22}-u_{12}^2 \,\rangle +
{qu\over p^2}\langle q_{11}u_{22}+u_{11}q_{22}\,\rangle  \\ 
+{q\over p}\langle w_{11}q_{22}+q_{11}w_{22}\,\rangle+
{u\over p}\langle w_{11}u_{22}+u_{11}w_{22}\,\rangle \nonumber
\label{dlt}
\ee
and for {\it conditional} mean values we have
\be 
\langle q_{11}q_{22}-q_{12}^2\rangle  = {\gamma^2\over 4}(q^2-1), \nonumber \\ 
\langle u_{11}u_{22}-u_{12}^2\rangle  = {\gamma^2\over 4}(q^2-1), \nonumber \\ 
\langle q_{11}u_{22}+u_{11}q_{22}\rangle =2{\gamma^2\over 4} qu, 		\\ 
\langle w_{11}q_{22}+q_{11}w_{22}\rangle =-{\gamma^2\over 4}{q\over p},\nonumber \\ 
\langle w_{11}u_{22}+u_{11}w_{22}\rangle =
-{\gamma^2\over 4}{u\over p},\nonumber 
\label{quw}
\ee
Finally
\be 
\langle \Delta(p)\,\rangle   = {\gamma^2\over 4}(p^2-3),\nonumber \\ 
G(p_t)=\frac{1}{2\pi r_c^2}(p_t^2-1)e^{-\frac{p_t^2}{2}}
\label{mdlt}
\ee

More cumbersome method for the calculation of the Genus was used in 
ref.~\cite{nn} where the calculations  have been done under  
incorrect assumptions about the correlations of $p_{ij}$. In spite of 
that, the final result obtained there remains true. 

All three MF's have very simple analytical forms and can
be used, together with other methods, to test the Gaussianity of CMB. 

As was noted in ref.~\cite{mw} and later discussed in refs.~\cite{nn} and 
\cite{ddnn} the condition $G(p_{pr}) =0$ defines the critical amplitude 
$p_{pr}$ for which the regions of high polarization percolate. According to 
eq. (\ref{mdlt}) this critical amplitude of polarization is 
$$p = p_{pr}=1$$
that was tested in simulations made in ref.~\cite{nn}. 

As has been mentioned above, $G(p_t)$ is equal to the number of 
isolated regions above $p_t$ minus the number of isolated regions below 
$p_t$. For $p_t\ll 1$ the percolation takes place, $A\approx 1$ and 
almost all map  
is occupied by the one region with  $p\geq p_t$. This means that 
only small isolated spots around the points $p = 0$ can be separated 
and the mean number density of such points, $N_0$, is
\be 
N_0 = 1-G(0) = \frac{1}{2\pi r_c^2}
\label{g0}
\ee

As we can see from eqs. (\ref{rc}), (\ref{gamma}) and  (\ref{dimless})
MF's have an universal form for any relation between the potentials $\Phi$ 
and $\Psi$.
Therefore, Minkowski Functionals for the CMB polarization field do not 
depend on the
type of perturbations and are determined by the statistics of the
signal only. To discriminate the scalar and pseudo-scalar perturbations 
and to estimate the parameter $\rho$ given by eq. (69),
a more detailed investigation of statistics of $Q_{ij}, U_{ij}$ and  $
p_{ij}$ needs to be performed. This implies, for example, direct 
estimates of some of the correlators (69) which would be a very 
difficult problem because of a finite resolution of the future
polarization maps.

\section{Summary and discussion}

The main goal of this paper is to present a critical discussion of 
the known methods of analysis
of statistical and geometrical properties of CMB polarization and to 
propose some new additional methods.

We have compared local and nonlocal descriptions of the polarization
with different expressions for the field functionals $S$ and $P$ (see
Sec. 3) in the case of noisy data. We conclude that the optimal
observational strategy strongly depends upon the properties of the 
noise. For a sufficiently rare sources of noise on the sky, a measurement
of local quantities $S$ and $P$ (\ref{sp}) which are obtained by 
differentiation of the Stokes parameters seems to be more favorable.
However an isotropic noise with zero average may be easier to suppress
by taking an average value over a part of the sky.

We established a classification of the singular points of the flux lines
of the eigen-vector $\vec n^{(+)}$ of the polarization (Stokes) matrix.
We have found that there are three possible types of singularities at the
points where polarization vanishes: saddles, comets, and beaks. They are
different from the singularities known in the standard theory of two 
dimensional vector fields. The former, as is well known, are saddles, knots,
and foci (even though the name "saddle" is the same in both cases the 
behavior of the flux lines are quite different). The reason for this 
difference is a square root singularity of the vector $\vec n^{(+)}$ at 
the point where polarization vanishes. 
We have compared our method which
was earlier presented in ref.~\cite{ddnn} and described in detail here,
with other methods~\cite{nn,geom1,geom2}. 

It is shown that in the case of Gaussian primordial fluctuations (as 
predicted by inflationary cosmology) the saddles make 50\% of all
singular points, beaks make 5.2\%, and comets make 44.8\%. Deviations 
from these numbers may signal deviations from a Gaussian nature of
perturbations or may indicate an existence of a non-Gaussian noise in
observational data. We realize that a measurement of polarization near
the points where it vanishes is a formidable observational problem.
But the patterns on polarization maps created by the flux lines of
$\vec n^{(+)}$ in the regions where polarization is non-zero is to a 
large extent determined by the types of nearest singularities. Thus
one may hope that measurements of polarization on the patches where it
is sufficiently large would permit to determine the singularity types.

Finally we applied Minkowski Functionals to the analysis of the statistic
of the geometrical quantities on the polarization maps.It is shown that
it provides a sensitive test of Gaussianity of CMB.

\bigskip
{\bf {Acknowledgments}}

This work was supported in part by Danmarks Grundforskningsfond through
its funding of the Theoretical Astrophysical Center (TAC), the Danish
Natural Science Research Council through grant No 9401635, and in part
by NSF-NATO fellowship (DGE-9710914) and by NSF EPSCOR program.
D.N. thanks M. Kamionkowski for helpful discussions.

\end{document}